\documentclass[preprintnumbers,amsmath,amssymbm,preprint]{revtex4}
\usepackage{epsfig}
\usepackage{graphicx}

\begin{document}
\title{Marginally bound
%spherical
(critical) geodesics of rapidly rotating black holes}
\author{Shahar Hod}
\address{The Ruppin Academic Center, Emeq Hefer 40250, Israel}
\address{ }
\address{The Hadassah Institute, Jerusalem 91010, Israel}
\date{\today}

\begin{abstract}
One of the most important geodesics in a black-hole spacetime is the
marginally bound spherical orbit. This critical geodesic represents
the innermost spherical orbit which is bound to the central black
hole. The radii $r_{\text{mb}}({\bar a})$ of the marginally bound
{\it equatorial} circular geodesics of rotating Kerr black holes
were found analytically by Bardeen {\it et. al.} more than four
decades ago (here $\bar a\equiv J/M^2$ is the dimensionless
angular-momentum of the black hole). On the other hand, no
closed-form formula exists in the literature for the radii of
generic ({\it non}-equatorial) marginally bound geodesics of the
rotating Kerr spacetime. In the present study we analyze the
critical (marginally bound) orbits of rapidly rotating Kerr black
holes. In particular, we derive a simple {\it analytical} formula
for the radii $r_{\text{mb}}(\bar a\simeq 1;\cos i)$ of the
marginally bound spherical orbits, where $\cos i$ is an effective
inclination angle (with respect to the black-hole equatorial plane)
of the geodesic. We find that the marginally bound spherical orbits
of rapidly-rotating black holes are characterized by a critical
inclination angle, $\cos i=\sqrt{{2/3}}$, above which the coordinate
radii of the geodesics approach the black-hole radius in the
extremal $\bar a\to1$ limit. It is shown that this critical
inclination angle signals a transition in the physical properties of
the orbits: in particular, it separates marginally bound spherical
geodesics which lie a finite proper distance from the black-hole
horizon from marginally bound geodesics which lie an infinite proper
distance from the horizon.
\end{abstract}
%\bigskip
\maketitle
%]

\section{Introduction}

The geodesic motions of test particles in the rotating Kerr
black-hole spacetime have attracted much attention since the
pioneering work of Carter \cite{Car}, see also
\cite{Bar2,Chan,Shap,CarC,Ryan4,Will,Mar,Hod} and references
therein. Among the various geodesics which characterize the
black-hole spacetime, the single most important family of geodesics
are the spherical orbits
--- orbits with constant coordinate radii on which test particles
can travel around the central black hole.

As emphasized in \cite{Bar2}, not all spherical orbits are bound to
the black hole. An unbound geodesic is characterized by $E/\mu>1$,
where $E$ and $\mu$ are the total (conserved) energy and rest mass
of the orbiting particle, respectively. Given an infinitesimal
outward perturbation, a particle on an unbound geodesic will escape
to infinity \cite{Bar2}. The unbound spherical orbits are separated
from the bound orbits by a critical geodesic called the {\it
marginally bound} spherical geodesic. This important geodesic is
characterized by a zero binding energy $E_{\text{binding}}\equiv
E(\infty)-E(r_{\text{mb}})=0$, or equivalently
\begin{equation}\label{Eq1}
E(r_{\text{mb}})=\mu\  .
\end{equation}
[The critical orbit so defined is sometimes referred to as the IBSO
(innermost bound spherical orbit) \cite{Ryan4,Will}].

The marginally bound spherical orbit is interesting from both an
astrophysical and theoretical points of view. In particular, this
critical geodesic plays an important role in the evolution and
dynamics of star clusters around supermassive galactic black holes
\cite{Will,Mar}. As emphasized in \cite{Bar2}, any parabolic orbit
which penetrates below the innermost bound spherical orbit must
plunge directly into the central black hole.

It is well-known that astrophysically realistic black holes
generally possess angular momentum. An astrophysically realistic
model of particle-dynamics in a black-hole spacetime should
therefore involve a non-spherical Kerr geometry \cite{Hod2000}. It
should be emphasized that, for rotating Kerr black holes, the
physical properties (in particular, the characteristic radii and
angular-momentum) of the marginally bound spherical orbits must be
computed {\it numerically}. The only known exceptions are the
co-rotating and counter-rotating marginally bound circular geodesics
in the equatorial plane of the Kerr black hole, in which case a
closed analytical formula [see Eq. (\ref{Eq20}) below] for the radii
of these special orbits has been given in \cite{Bar2}.

To the best of our knowledge, no closed-form formula exists in the
literature for the radii of generic ({\it non}-equatorial)
marginally bound geodesics which characterize the rotating Kerr
black-hole spacetime. As we shall show below, for rapidly-rotating
black holes one can obtain a simple and compact {\it analytical}
formula for the characteristic radii of the marginally bound
spherical geodesics. In particular, we shall show below that
rapidly-rotating Kerr black holes are characterized by a significant
fraction of marginally bound geodesics whose radii approach the
black-hole radius in the near-extremal limit.

The rest of the paper is devoted to the investigation of the
physical properties of non-equatorial marginally bound spherical
geodesics of the rotating Kerr black-hole spacetime. In Sec. II we
describe the dynamical equations which determine the geodesics of
test particles in the rotating Kerr spacetime. In Sec. III we derive
the characteristic equation which determines the radii of the
generic (non-equatorial) marginally bound spherical orbits. In Sec.
IV we focus on near-extremal black holes and analyze the marginally
bound geodesics of these rapidly-rotating black holes. We conclude
in Sec. V with a brief summary of our results.

\section{Description of the system}

We shall analyze the spherical geodesics of test particles in the
spacetime of a rapidly-rotating Kerr black hole. In Boyer-Lindquist
coordinates the metric is given by (we use gravitational units in
which $G=c=1$) \cite{Chan,Kerr}
\begin{eqnarray}\label{Eq2}
ds^2=-\Big(1-{{2Mr}\over{\rho^2}}\Big)dt^2-{{4Mar\sin^2\theta}\over{\rho^2}}dt
d\phi+{{\rho^2}\over{\Delta}}dr^2 \nonumber \\
+\rho^2d\theta^2+\Big(r^2+a^2+{{2Ma^2r\sin^2\theta}\over{\rho^2}}\Big)\sin^2\theta
d\phi^2,
\end{eqnarray}
where $M$ and $a$ are the mass and angular momentum per unit mass of
the black hole, respectively. Here $\Delta\equiv r^2-2Mr+a^2$ and
$\rho\equiv r^2+a^2\cos^2\theta$. The black-hole (event and inner)
horizons are located at the zeroes of $\Delta$:
\begin{equation}\label{Eq3}
r_{\pm}=M\pm(M^2-a^2)^{1/2}\  .
\end{equation}

The general geodesics of test particles in the Kerr black-hole
spacetime are characterized by four constants of the motion
\cite{Car}. In terms of the covariant Boyer-Lindquist components of
the $4$-momentum, these conserved quantities are \cite{Bar2}:
\begin{equation}\label{Eq4}
E\equiv -p_t=\text{total energy}\  ,
\end{equation}
\begin{equation}\label{Eq5}
L_z\equiv p_{\phi}=\text{component of angular momentum parallel to
the symmetry axis}\  ,
\end{equation}
\begin{equation}\label{Eq6}
Q\equiv
p^2_{\theta}+\cos^2\theta[a^2(\mu^2-p^2_t)+p^2_{\phi}/\sin^2\theta]\
,
\end{equation}
and
\begin{equation}\label{Eq7}
\mu=\text{the rest mass of the particle}\ .
\end{equation}

The geodesics in the black-hole spacetime are governed by the
following set of equations \cite{Bar2}
\begin{equation}\label{Eq8}
\rho {{dr}\over{d\lambda}}=\pm\sqrt{V_r}\  ,
\end{equation}
\begin{equation}\label{Eq9}
\rho {{d\theta}\over{d\lambda}}=\pm\sqrt{V_{\theta}}\  ,
\end{equation}
\begin{equation}\label{Eq10}
\rho{{d\phi}\over{d\lambda}}=(L_z/\sin^2\theta-aE)+aT/\Delta\  ,
\end{equation}
\begin{equation}\label{Eq11}
\rho{{dt}\over{d\lambda}}=a(L_z-aE\sin^2\theta)+(r^2+a^2)T/\Delta\ ,
\end{equation}
where $\lambda$ is related to the particle's proper time by
$\lambda=\tau/\mu$. Here
\begin{equation}\label{Eq12}
T\equiv E(r^2+a^2)-L_za\  ,
\end{equation}
\begin{equation}\label{Eq13}
V_r\equiv T^2-\Delta[\mu^2r^2+(L_z-aE)^2+Q]\  ,
\end{equation}
\begin{equation}\label{Eq14}
V_{\theta}\equiv Q-\cos^2\theta[a^2(\mu^2-E^2)+L^2_z/\sin^2\theta]\
.
\end{equation}
The effective potentials $V_r$ and $V_{\theta}$ determine the
orbital motions in the $r$ and $\theta$ directions, respectively
\cite{Ryan4,Noteref6}.

Circular equatorial orbits are characterized by $Q=0$ \cite{Bar2}.
It is convenient to use an effective inclination angle $i$ to
quantify the deviation of a generic (non-equatorial) orbit from the
equatorial plane of the black hole. The effective inclination angle
is defined by \cite{Ryan1,Ryan2,Ryan3,Ryan4}
\begin{equation}\label{Eq15}
\cos i\equiv {{L_z}\over{L}}\  ,
\end{equation}
where
\begin{equation}\label{Eq16}
L\equiv\sqrt{L^2_z+Q}\  .
\end{equation}
Note that $L$ and $i$ are constants of the motion. For spherical
black-hole spacetimes (with $a=0$), $L$ is the total angular
momentum of the orbiting particle \cite{Bar2}. The extensively
studied equatorial orbits are characterized by $\cos^2 i=1$, where
$\cos i=+1$$/$$-1$ correspond to co-rotating/counter-rotating
geodesics, respectively.

\section{Marginally bound spherical geodesics of the black-hole spacetime}

Spherical geodesics in the black-hole spacetime are characterized by
the two conditions \cite{Chan,Bar2}
\begin{equation}\label{Eq17}
V_r=0\ \ \ \text{and}\ \ \ V'_r=0\  .
\end{equation}
Substituting (\ref{Eq13}) into (\ref{Eq17}) and using the condition
$E=\mu$ [see Eq. (\ref{Eq1})] for the marginally bound orbits, one
obtains the dimensionless angular-momentum-to-energy ratio
\begin{equation}\label{Eq18}
{{L}\over{M\mu}}=\sqrt{{{4r^3}\over{M(r^2-a^2\sin^2i)}}}\
\end{equation}
and the characteristic equation
\begin{eqnarray}\label{Eq19}
r^4-4Mr^3-a^2(1-3\sin^2i)r^2+a^4\sin^2i+4a\cos
i\sqrt{Mr^5-Ma^2r^3\sin^2i}=0\
\end{eqnarray}
for the radii $r_{\text{mb}}(M,a;\cos i)$ of the marginally bound
spherical orbits.

The exact (analytical) solution of the characteristic equation
(\ref{Eq19}) is only known for the simple case of circular geodesics
in the equatorial plane of the black hole. As mentioned above, these
special orbits are characterized by $\cos i=\pm 1$, where the upper
sign corresponds to the co-rotating circular orbit while the lower
sign corresponds to the counter-rotating circular orbit. In this
simple case one finds \cite{Bar2}
\begin{equation}\label{Eq20}
r_{\text{mb}}(M,a;\cos i=\pm 1)=2M\mp a +2M^{1/2}(M\mp a)^{1/2}\  .
\end{equation}

To the best of our knowledge, no closed-form formula exists in the
literature for the radii $r_{\text{mb}}(M,a;\cos i)$ of generic
(non-equatorial, $\cos i\neq \pm 1$) marginally bound spherical
geodesics of the Kerr spacetime. It is worth mentioning that Will
\cite{Will} also studied the marginally bound spherical geodesics of
the Kerr spacetime. In particular, Ref. \cite{Will} provides an {\it
expansion} of the physical quantities in powers of the dimensionless
ratio $a/M$. As emphasized in \cite{Will}, this expansion in powers
of $a/M$ works extremely well (to better than $0.5\%$) in the regime
$0\leq a/M\leq 0.9$, but is less accurate in the regime of
rapidly-rotating (near-extremal) black holes. [Note that \cite{Will}
also provides results for the marginally bound geodesics of exactly
extremal (with $a$ {\it exactly} equals $M$) black holes.] In the
present paper we shall perform an analysis which is complementary to
the one presented in \cite{Will}. In particular, we shall provide an
alternative expansion of the physical quantities in the small
parameter $(r_+-M)/M\ll 1$. This expansion is valid for
rapidly-rotating (but {\it not} necessarily extremal) black holes
(see \cite{Will} for the exactly extremal $a=M$ case). We shall
derive a compact analytical expression for the characteristic radii
of these critical (marginally-bound) orbits in the spacetime of
rapidly-rotating black holes. In particular, we shall show that the
dependence of $r_{\text{mb}}(a\simeq M;\cos i)$ on the effective
inclination angle of the orbit exhibits an interesting ``phase
transition" at the critical inclination angle $\cos i=\sqrt{{2/3}}$.

Before proceeding further, it is worth emphasizing that similar
phase transitions (with respect to the effective inclination angle)
are also found for other geodesics, including the stable circular
orbits \cite{Wilkins,Rufn,Hugn1,Hugn2,Bara}. In particular, Wilkins
\cite{Wilkins} studied the spherical geodesics of a maximally
rotating ($a=M$) black hole. It was found in \cite{Wilkins} that,
for exactly extremal $a=M$ black hole, there is a very interesting
family of orbits whose coordinate radius coincides with the
coordinate radius of the (extremal) black-hole horizon, $r=r_+=M$.
Johnston and Ruffini \cite{Rufn} extended the results of
\cite{Wilkins} to the case of charged Kerr-Newman black holes. Our
analysis extends the results of \cite{Wilkins} to the regime of
rapidly-rotating (but {\it not} necessarily extremal) black holes.
%In addition, Hughes \cite{Hugn1,Hugn2} identified an interesting
%family of spherical geodesics of near-extremal black holes with the
%unusual behavior $\partial L_z / \partial i >0$.
It is worth mentioning that an analysis analogous to the one
presented here reveals that the innermost stable circular orbit
(ISCO) is characterized by a similar phase transition which, for
near-extremal black holes, occurs at the effective inclination angle
$\cos i=\sqrt{{4/5}}$. In addition, the null spherical geodesics
\cite{Chan,Teo,Yang} of near-extremal Kerr black holes are also
characterized by an analogous phase transition which occurs at the
effective inclination angle $\cos i=\sqrt{{4/7}}$ \cite{Hod}.

\section{Rapidly-rotating (near-extremal) black holes}

For rapidly-rotating (near-extremal) black holes it is convenient to
define
\begin{equation}\label{Eq21}
r_{\pm}\equiv M(1\pm\epsilon)\ \ \ \text{and} \ \ \
r_{\text{mb}}\equiv M(1+\delta_{\text{mb}})\  ,
\end{equation}
where $\epsilon,\delta_{\text{mb}}\ll 1$. From (\ref{Eq21}) one
finds
\begin{equation}\label{Eq22}
a=M[1-\epsilon^2/2+O(\epsilon^4)]\  .
\end{equation}
Substituting (\ref{Eq21}) and (\ref{Eq22}) into the characteristic
equation (\ref{Eq19}), one finds
\begin{eqnarray}\label{Eq23}
&(\cos^2i-2)(3\cos^2i-2)\delta^2+[(\cos^6i-18\cos^4i+20\cos^2i-8)/2\cos^2i]\delta^3
\nonumber \\
& +2\cos^2i(2-\cos^2i)\epsilon^2+O(\delta^4,\epsilon^4)=0\
\end{eqnarray}
for the marginally bound orbits of rapidly-rotating black holes. The
qualitative behavior of $\delta_{\text{mb}}(\epsilon;\cos i)$ in the
near-extremal $\epsilon\to 0$ limit depends on whether the
coefficient of the $O(\delta^2)$ term in Eq. (\ref{Eq23}) is
positive, negative, or zero. Note that this coefficient vanishes at
the critical inclination angle
\begin{equation}\label{Eq24}
\xi\equiv\cos i-\sqrt{{2/3}}=0\  .
\end{equation}
The solution of the characteristic equation (\ref{Eq23}) is given by
\cite{Notesol}
\begin{equation}\label{Eq25}
\delta_{\text{mb}}(\epsilon;\cos i)=
\begin{cases}
\sqrt{{{2\cos^2i}\over{3\cos^2i-2}}}\ \epsilon+O(\epsilon^2/\xi^2)\
&
\text{ for }\ \ \cos i-\sqrt{{2/3}}\gg \epsilon^{2/3}\ ; \\
\epsilon^{2/3}+O(\xi) & \text{ for }\ \ -\epsilon^{2/3}\ll \cos i-\sqrt{{2/3}}\ll \epsilon^{2/3}\ ; \\
{9\over{\sqrt{6}}}(\sqrt{{2/3}}-\cos i)+O(\epsilon^2/\xi^2) &
\text{ for }\ \ \epsilon^{2/3}\ll\sqrt{{2/3}}-\cos i\ll 1 \ , \\
\end{cases}
\end{equation}
From (\ref{Eq25}) one learns that the solution
$\delta_{\text{mb}}(\epsilon;\cos i)$ of the characteristic equation
(\ref{Eq23}) exhibits a ``phase transition" [from a
$\delta_{\text{mb}}(\epsilon\to 0)\to 0$ behavior to a finite
$\delta_{\text{mb}}(\epsilon\to 0)$ behavior], which occurs in the
extremal $\epsilon\to 0$ limit at the critical inclination angle
$\cos i=\sqrt{{2/3}}$.

%In figure \ref{Fig1} we display the effective radial potential $V_r$
%[see Eqs. (\ref{Eq1}), (\ref{Eq12}), (\ref{Eq13}), (\ref{Eq15}), and
%(\ref{Eq18})] for various values of the effective inclination angle.

We shall now show that the critical inclination angle, $\cos
i=\sqrt{{2/3}}$, separates marginally bound orbits which are
characterized by finite proper distances to the black-hole horizon
from marginally bound orbits which are characterized by infinite
proper distances to the horizon. The proper radial distance between
the black-hole horizon [at $r_+=M(1+\epsilon)$] and the intersection
point of the marginally bound orbit [of radius
$r_{\text{mb}}=M(1+\delta_{\text{mb}})$] with the equatorial plane
of the black hole is given by (we emphasize that we use here the
Boyer-Lindquist coordinates) \cite{Ted}
\begin{equation}\label{Eq26}
\Delta\ell=M[\sqrt{\delta^2-\epsilon^2}+\ln\big(\delta+\sqrt{\delta^2-\epsilon^2}\big)-\ln\epsilon]\
.
\end{equation}
Substituting $\delta_{\text{mb}}(\epsilon;\cos i)$ from Eq.
(\ref{Eq25}) into Eq. (\ref{Eq26}), one finds
\begin{equation}\label{Eq27}
\Delta\ell(\epsilon\to 0)=M\times
\begin{cases}
\ln\Big({{\sqrt{2}\cos
i+\sqrt{2-\cos^2i}}\over{\sqrt{3\cos^2i-2}}}\Big)+O(\epsilon)\ &
\text{ for }\ \ \cos i-\sqrt{{2/3}}\gg \epsilon^{2/3}\ ; \\
-{1\over 3}\ln\epsilon+O(1) & \text{ for }\ \ -\epsilon^{2/3}\ll \cos i-\sqrt{{2/3}}\ll \epsilon^{2/3}\ ; \\
-\ln\epsilon+O(\ln\xi) &
\text{ for }\ \ \epsilon^{2/3}\ll\sqrt{{2/3}}-\cos i\ll 1 \ , \\
\end{cases}
\end{equation}
in the extremal $\epsilon\to 0$ limit. One therefore concludes that
the critical inclination angle, which occurs at $\cos
i=\sqrt{{2/3}}$ in the near-extremal limit, signals a transition
from {\it finite} to {\it infinite} proper distances of the
marginally bound orbits from the black-hole horizon \cite{NoteBar}.

Each marginally bound orbit is bounded in some strip
$[\theta^{-},\theta^{+}]$ of the polar angle $\theta$, where
$\theta^{-}=\pi-\theta^{+}$. The two polar turning-points are
determined from the requirement
\begin{equation}\label{Eq28}
V_{\theta}(\theta^{\pm})=0\  ,
\end{equation}
see Eq. (\ref{Eq14}). We shall now evaluate the polar boundaries
$\{\theta^-,\theta^+\}$ of the marginally bound near-horizon
geodesics (the marginally bound orbits which are characterized by
finite proper distances to the black-hole horizon in the
near-extremal limit). Taking cognizance of Eqs. (\ref{Eq18}),
(\ref{Eq21}), and (\ref{Eq25}), one finds the dimensionless ratio
\begin{equation}\label{Eq29}
{L\over {M\mu}}={{2}\over{\cos
i}}\Big[1+\sqrt{{{3\cos^2i-2}\over{2\cos^2i}}}\
\epsilon+O(\epsilon^2)\Big]
\end{equation}
for the marginally bound spherical orbits in the regime $\cos
i-\sqrt{{2/3}}\gg \epsilon^{2/3}$ \cite{Notewil}. Substituting
(\ref{Eq29}) into (\ref{Eq28}) [see Eq. (\ref{Eq14}) for
$V_{\theta}$], one finds that the polar turning-points (which
characterize the maximal polar-deviation of a near-horizon geodesic
from the equatorial plane of the black hole) are given by the simple
relation \cite{Noteinc}
\begin{equation}\label{Eq30}
\cos^2\theta^{\pm}=\sin^2i\  .
\end{equation}
For the equatorial circular orbit with $\cos i=1$ one finds
$\cos\theta^{\pm}=0$ ($\theta^{\pm}={\pi \over 2}$) as expected. For
the critical marginally bound geodesic with $\cos i=\sqrt{{2/3}}$
[see Eq. (\ref{Eq24})] one finds from (\ref{Eq30})
\begin{equation}\label{Eq31}
\cos\theta^{\pm}=\pm{{1}\over{\sqrt{3}}}\  .
\end{equation}

\section{Summary}

We have studied {\it analytically} the critical (marginally bound)
spherical geodesics of rapidly-rotating Kerr black holes. While most
former studies have focused on
%(the relatively simple case of)
circular orbits in the equatorial plane of the black hole
\cite{Bar2}, in the present paper we have considered generic ({\it
non}-equatorial) marginally bound geodesics. In particular, we have
derived an analytical expression [see Eq. (\ref{Eq25})] for the
radii $r_{\text{mb}}(a\simeq M;\cos i)$ of the critical (marginally
bound) spherical geodesics which characterize the rapidly-rotating
Kerr black-hole spacetime.

The analytical formula (\ref{Eq25}) reveals that the marginally
bound spherical geodesics of rapidly-rotating black holes are
characterized by a critical inclination angle, $\cos
i=\sqrt{{2/3}}$, above which the coordinate radii of the orbits
approach the black-hole radius in the extremal limit. We have proved
that this critical inclination angle signals a transition in the
physical properties of the marginally bound orbits: in particular,
it separates marginally bound geodesics which lie a finite proper
distance from the black-hole horizon from marginally bound geodesics
which lie an infinite proper distance from the horizon.

\bigskip
\noindent
{\bf ACKNOWLEDGMENTS}
\bigskip

This research is supported by the Carmel Science Foundation. I thank
Yael Oren, Arbel M. Ongo and Ayelet B. Lata for helpful discussions.

%\newpage

%\newpage

%\input{epsf}
%\begin{figure}[h]
%  \begin{center}
%    \epsfxsize=8.5cm \epsffile{Fig1.eps}
%  \end{center}
%  \caption{The effective radial potential $V_r$ for an extremal Kerr black hole and for various values
%  of the effective inclination angle.}
%  \label{Fig1}
%\end{figure}


\begin{thebibliography}{99}
\bibitem{Notesol} In obtaining the solution (\ref{Eq25}) for $\delta_{\text{mb}}(\epsilon;\cos i)$ we
have used the observation that, depending of the relative magnitudes
of $\epsilon$ and $\xi$, one of the three terms in the
characteristic equation (\ref{Eq23}) can be neglected as compared to
the other two terms. It should be emphasized that one of the two
remaining (non-negligible) terms must be positive while the other
non-negligible term must be negative. Specifically, there are three
distinct regimes: (1) For $\xi\gg \epsilon^{2/3}$ the $O(\delta^3)$
term is negligible as compared to the other two terms in Eq.
(\ref{Eq23}). (2) For $-\epsilon^{2/3}\ll \xi\ll \epsilon^{2/3}$ the
$O(\xi\delta^2)$ term is negligible as compared to the other two
terms in Eq. (\ref{Eq23}). (3) For $\xi\ll -\epsilon^{2/3}$ the
$O(\epsilon^2)$ term is negligible as compared to the other two
terms in Eq. (\ref{Eq23}).
\bibitem{NoteBar} Note that one finds $\Delta\ell=M\ln(1+\sqrt{2})$ for the prograde equatorial geodesic with
$\cos i=1$, in accord with the result of \cite{Bar2}.
\bibitem{Notewil} Note that one finds $L/M\mu=2/\cos i$ for extremal black holes
with $\epsilon\equiv 0$, in accord with the result of \cite{Will}.
\bibitem{Noteinc} It is worth emphasizing again that the effective inclination angle $i$ is a
{\it constant} of the motion, while the polar angle $\theta$ varies
along the trajectory within the interval $[\theta^-,\theta^+]$,
where the two boundaries are determined by the relation
(\ref{Eq30}).
\end{thebibliography}
\end{document}